\documentclass[10pt,letterpaper,pra,twocolumn,superscriptaddress]{revtex4-1}
\usepackage{graphicx,color}
\usepackage{bm}
\usepackage{float}
\usepackage{ulem}
\usepackage{amsmath}
\usepackage[utf8]{inputenc}
\usepackage[colorlinks=true,linkcolor=blue,citecolor=magenta,urlcolor=blue]{hyperref}

\hypersetup{colorlinks=true,linkcolor=blue,citecolor=blue,urlcolor=blue}

\begin{document}

\title{High-accuracy adaptive quantum tomography for high-dimensional quantum systems}

\author{L. Pereira}
\affiliation{Instituto de F\'{\i}sica Fundamental IFF-CSIC, Calle Serrano 113b, Madrid 28006, Spain.\\}
\author{D. Mart\'inez}
\affiliation{Millennium Institute for Research in Optics, Universidad de Concepci\'on, Concepci\'on, Chile}
\affiliation{Departamento de F\'isica, Universidad de Concepci\'on, 160-C Concepci\'on, Chile}
\author{G. Ca\~nas}
\affiliation{Millennium Institute for Research in Optics, Universidad de Concepci\'on, Concepci\'on, Chile}
\affiliation{Departamento de F\'isica, Universidad del B\'io-B\'io, Collao 1202, Casilla 5C, Concepci\'on, Chile}
\author{E.~S.~G\'omez}
\author{S.~P.~Walborn}
\author{G. Lima}
\author{A. Delgado}
\affiliation{Millennium Institute for Research in Optics, Universidad de Concepci\'on, Concepci\'on, Chile}
\affiliation{Departamento de F\'isica, Universidad de Concepci\'on, 160-C Concepci\'on, Chile}

\date{\today}

\begin{abstract}
The accuracy of estimating $d$-dimensional quantum states is limited by the Gill-Massar bound. It can be saturated in the qubit ($d=2$) scenario using adaptive standard quantum tomography. In higher dimensions, however, this is not the case and the accuracy achievable with adaptive quantum tomography quickly deteriorates with increasing $d$. Moreover, it is not known whether or not the Gill-Massar bound can be reached for an arbitrary $d$. To overcome this limitation, we introduce an adaptive tomographic method that is characterized by a precision that is better than half that of the Gill-Massar bound for any finite dimension.  This provides a new achievable accuracy limit for quantum state estimation. We demonstrate the high-accuracy of our method by estimating the state of 10-dimensional quantum systems. With the advent of new technologies capable of high-dimensional quantum information processing, our results become critically relevant as state reconstruction is an essential tool for certifying the proper operation of quantum devices.
\end{abstract}

\maketitle

\maketitle
{\it Introduction.---} High-dimensional quantum systems (qudits) offer several advantages over two-dimensional quantum systems for information processing tasks. For instance, the performance of some practical protocols of quantum communication and computation is enhanced when qudits are employed \cite{Cerf,Armin_2017,Armin_2018,Araujo_2014,Taddei_2020}. These also lead to improvements in entanglement-based fundamental studies since, in this case, Bell inequalities exhibit a higher robustness against noise and tolerate lower detection efficiencies \cite{bell,Brunner_2010,Zeilinger_2000,cglmp}. However, as the dimension of quantum systems increases, experimental realizations become increasingly challenging \cite{QRAC1024,Dada2011,Krenn2014,Blatt2011,Pan2016,Guo_2020,GuixReview_2019}, resulting in reduced qudit generation rates that lead to smaller ensembles of identical and independently prepared copies. This greatly decreases the achievable accuracy of quantum state estimation techniques.

Here, we study the problem of estimating quantum states of qudits with the highest possible accuracy taking into account the dimension of the quantum systems as well as the size of the generated ensembles. In particular, we introduce a method for high-accuracy quantum tomography (HAQT) that is based on the strategy of adaptive projective measurements. Measurement adaptation \cite{Straupe} has been suggested as a means to increase the estimation accuracy of tomographic schemes \cite{Okamoto,Mahler,Li,Zhu,Utreras-Alarcon,Zambrano}. In the case of $d=2$, adaptive quantum tomography (AQT) \cite{Huo} reaches the optimal estimation accuracy given by the Gill-Massar lower bound \cite{Gill-Massar}. In higher dimensions this is not the case and AQT rapidly departs from optimality \cite{Pereira}. Unlike previous results, we can show that our method is characterized by an estimation accuracy that is at most twice the optimal estimation accuracy. This holds for quantum systems of arbitrary finite dimension. Since it is not known whether or not the Gill-Massar lower bound can be attained, our result establishes an upper bound for the achievable estimation accuracy of qudits. We demonstrate the high accuracy of our tomographic method by estimating the state of photonic path qudits in dimension $d=10$, which are defined in terms of the transverse momentum of single photons transmitted through programmable spatial light modulators \cite{GLima01,GLima03}.

{\it High-accuracy quantum tomography.---} Quantum tomography of an unknown fixed state $\rho$ consists of acquiring enough information about it through measurements to construct an estimate $\tilde \rho$ of it. The accuracy of a tomographic method can be studied via the infidelity
\begin{equation}
I(\rho,\tilde\rho)=1-Tr^2(\sqrt{\sqrt{\rho}\tilde\rho\sqrt{\rho}}),
\label{Infidelity}
\end{equation}
which is a well-known distance between quantum states \cite{Jozsa}. The infidelity $I(\rho,\tilde{\rho})$ is zero for perfect state estimation ($\tilde{\rho}=\rho$) and its inverse can be identified with the sample size required to reach a given accuracy \cite{Mahler}. Experimentally, measurements are made on finite sets of $N$ equally prepared copies of the unknown state, and thus different runs of the same tomographic method can generate different estimates due to statistical fluctuations. Thereby, quantum tomography does not lead to a single estimate, but to a distribution of estimates defined by some probability density function $f(\tilde\rho)$. Therefore, in order to study the precision of the tomography, we use the mean infidelity defined by
\begin{equation}
\bar I(\rho)=\int I(\rho,\tilde\rho)f(\tilde\rho)d\tilde\rho.
\label{Mean-Infidelity}
\end{equation}
For a pair of states infinitesimally close to each other and with the same rank, the mean infidelity can be related to the mean squared Bures distance \cite{Hubner}, that is, $\bar I(\rho)=Tr ({\cal C}(\rho){\cal J})/4$, where $\cal{C}$ is the covariance matrix ${\cal C}_{j,k}(\rho)=\int(S_j-\tilde S_j)(S_k-\tilde S_k)f(\tilde \rho)d\tilde\rho$, with $S_j$ ($\tilde S_j$) parameters defining the state $\rho$ ($\tilde\rho$), and
\begin{equation}
\mathcal{J}_{j,k} = Tr[\rho(L_jL_k-L_kL_j) ]/2,
\end{equation}
the Quantum Fisher Information Matrix (QFIM), with $L_j$ the symmetric logarithmic derivatives, which are implicitly defined by $\partial\rho/\partial S_j=(\rho L_j+L_j\rho)/2$. Clearly, to increase the accuracy we have to reduce the Covariance Matrix. However, the covariance of an estimator cannot be arbitrarily reduced since Cramer-Rao inequalities \cite{Paris} ${\cal C}\ge{\cal I}^{-1}\ge{\cal J}^{-1}$ establish a fundamental lower bound for the Covariance Matrix. Here, ${\cal I}$ is the Classical Fisher Information Matrix (CFIM), defined by
\begin{equation}
{\cal I}_{j,k}=\sum_m \frac{1}{p_m}\frac{\partial p_m}{\partial S_j}\frac{\partial p_m}{\partial S_k}.
\label{CFisherInfo}
\end{equation}
In one-parameter estimation, the Cramer-Rao lower bound can be attained by measuring the symmetric logarithmic derivative. In the multi-parameter case, the estimation by separable measurements, that is, by measurements realized on each individual copy of the state, the quantum Cramer-Rao bound can not be attained and the fundamental bound for the mean-square Bures distance is given by the Gill-Massar lower bound \cite{Gill-Massar}
\begin{equation}
\bar I(\rho)\ge \bar I^{(opt)}=\frac{(d^2-1)(d+1)}{4N}.
\label{Gill-Massar-Bound}
\end{equation}
An estimate is considered to be optimal if it reaches the Gill-Massar lower bound, that is, the classical Cramer-Rao bound ${\cal C}^{-1}={\cal I}^{(opt)}$ and fulfills the condition \cite{Huo}
\begin{equation}
{\cal I}^{(opt)}=\frac{1}{d+1}{\cal J}.
\label{conditionIopt}
\end{equation}
Adaptive quantum tomography approaches the bound $\bar I^{(opt)}$. This method employs Standard Quantum Tomography (SQT), which is based on the measurement of the $d^2-1$ generalized Gell-Mann matrices, on a sub-ensemble of size $N_0$ to generate an estimate $\tilde\rho_0$. The eigenbasis of $\tilde\rho_0$ is subsequently used to represent the Gell-Mann matrices in a new stage of standard quantum tomography on an ensemble of size $N-N_0$. In the case of $d=2$, adaptive quantum tomography saturates the Gill-Massar lower bound \cite{Huo}. In higher dimensions, the mean infidelity generated by the adaptive procedure behaves as $O(1/N)$ for all quantum states. However, the mean infidelity departs from the Gill-Massar lower bound as the dimension increases. In the following we introduce an adaptive tomographic scheme built upon a set of bases such that the achieved mean infidelity $\bar I(\rho)$ is upper bounded by $2\bar I^{(opt)}$ for all dimension $d$, that is, $\bar I^{(opt)}\le\bar I(\rho)\le 2\bar I^{(opt)}$.

Our main aim is to provide a high accuracy estimate of the parameters $S_j$ that characterize the unknown state $\rho$. Let us start by considering a previously known state $\rho_0$ close to $\rho$ characterized by means of the parameters $S_j^0$. Thereby, these states are related through,
\begin{equation}
\rho=\rho_0+\frac{1}{2}\sum_{j=1}^{d^2-1}(S_j - S_j^0)\sigma_j,
\label{rho}
\end{equation}
where the $d^2-1$ hermitian, traceless operators $\sigma_j$ are the $d$-dimensional Gell-Mann operators. Since the state $\rho_0=\sum_{k=0}^{d-1}\lambda_k|k\rangle\langle k |$ is known, we can employ its eigenbasis to represent the Gell-Mann operators. Thus, the $d-1$ diagonal operators are
\begin{equation}
\sigma_k^D=\sqrt{\frac{2}{k(k+1)}}\left(\sum_{j=0}^{k-1}|j\rangle\langle j|-k|k\rangle\langle k |\right),
\end{equation}
with $k=0,\dots,d-1$, and the $d(d-1)$ non-diagonal operators $\sigma^A_{\alpha,j,k}$ are
\begin{equation}
\sigma^A_{\alpha,j,k}=i^\alpha\left(|j\rangle\langle k|+(-1)^{\alpha}|k\rangle\langle j|\right),
\end{equation}
with $\alpha=0,1$ and $0\le j<k\le d-1$. SQT is based on measurements of the Gell-Mann operators, or equivalently, on projective measurements on the vectors $\{|j\rangle \}$ and $\{ |\pm_{j,k}^{\alpha}\rangle = ( |j\rangle\pm i^{\alpha}|k\rangle )/\sqrt{2} \}$, which are eigenstates of $\sigma_k^D$ and the non-null eigenstates of $\sigma^A_{\alpha,j,k}$, respectively. This can be done by measuring each Gell-Mann operator independently. However, many Gell-Mann operators are simultaneously diagonalizable, so that we can obtain the same information with fewer observables. Next, we present a minimal set of bases that group the eigenstates of all the Gell-Mann operators. For $d$ odd, there are $2d$ bases given by
\begin{equation}
{\cal B}_{k+d\alpha}=\{|k\rangle, |\pm^\alpha_{k-\nu, k+\nu}\rangle, |\pm^\alpha_{D+k-\mu, D+k+1+\mu}\rangle\},
\label{Base-Odd}
\end{equation}
where $D=(d-1)/2, 0\le k\le d-1, 0\le\mu<|k-D|$ and $0<\nu\le D-|k-D|$ and operations on the subindexes are carried out $\mod(d)$. For $d$ even, there are $2d-1$ bases
\begin{eqnarray}
{\cal B}_{k+(d-1)\alpha}&=&\{|\pm_{k, d-1}^\alpha\rangle, |\pm^\alpha_{k-\nu, k+\nu}\rangle,
\nonumber\\
&&\quad|\pm^\alpha_{D+k-\mu, D+k+1+\mu}\rangle\},
\\
{\cal B}_{2d-2}&=&\{|0\rangle,\dots,|d-1\rangle\},
\end{eqnarray}
where $D=(d-2)/2, 0\le k\le d-2, 0\le\mu\le|k-D|$, $0<\nu\le D-|k-D|$, and operations on the subindexes are carried out $\mod(d-1)$.
\begin{figure*}[t]
	\centering
	\includegraphics[width=0.78 \textwidth]{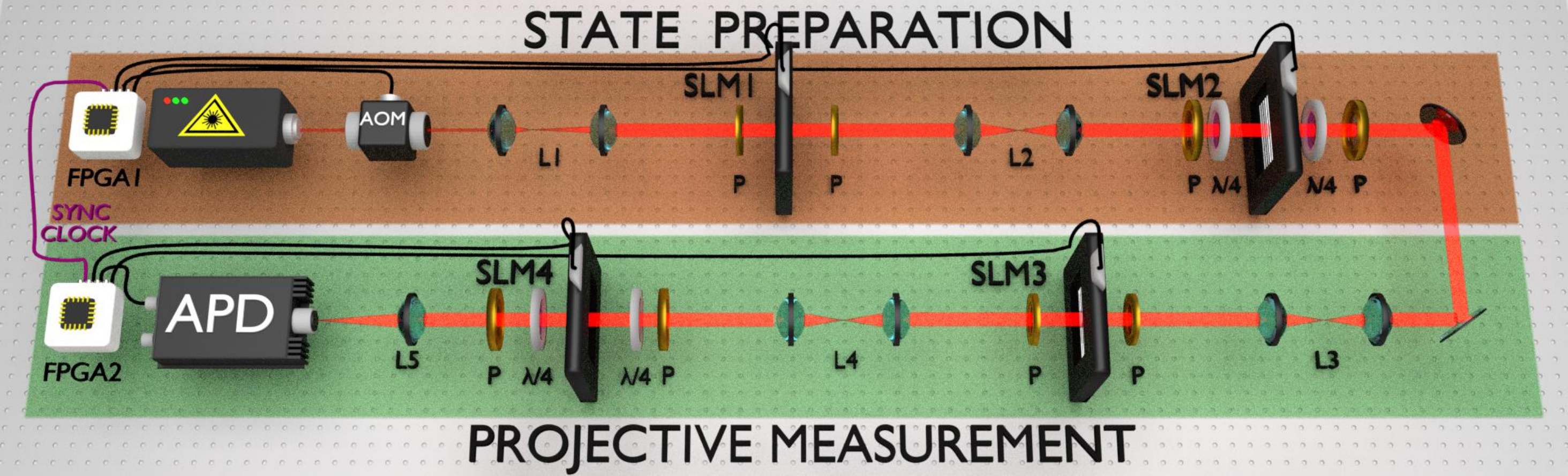}
	\caption{Experimental setup. The state preparation (SP) block consists of a weak coherent state source and a pair of SLMs to generate path qudit states in dimension $d=10$. The projective measurement (PM) block employs another pair of SLMs and an APD to measure the incoming path qudit state into the different projections required by our tomographic method. The experimental setup is automatically controlled by two FPGA electronic units, each of them located in each block. }
	 \label{fig:setup}
\end{figure*}
The number of bases can be resumed as $M_d = 2d-1+[d]$, where $[d] \equiv d \mod 2$. Thereby, to estimate $\rho$ our method measures each basis ${\cal B}_j$ on an ensemble of size $N/(2d-1+[d])$, instead of $N/(d^2-1)$ for all Gell-Mann operators. However a state $\rho_0$ close to $\rho$ is not always known. To overcome this limitation, an adaptive strategy can be used. A preliminary tomography is performed on sample of size $N_0<N$ using the measurements $\mathcal{B}_j$, but on a different basis, for example, the computational basis. From this tomography, we obtain a preliminary estimate $\tilde\rho_0$, which can be used as $\rho_0$. Subsequently, the tomographic method using the bases $\mathcal{B} _j$ is performed on the remaining sample $N-N_0$.

For study the accuracy of the method, we have to calculate the CFIM and the QFIM. This can be done using the fact that $\rho_0$ and $\rho$ are infinitesimally close. Considering a block representation of both matrices
\begin{equation}
	\mathcal{I}=\begin{pmatrix}
					\mathcal{I}^A& \mathcal{I}^{AD}\\ \mathcal{I}^{AD}&\mathcal{I}^D
				\end{pmatrix}, \quad
	\mathcal{J}=\begin{pmatrix}
					\mathcal{J}^A& \mathcal{J}^{AD}\\ \mathcal{J}^{AD}&\mathcal{J}^D
				\end{pmatrix},
\end{equation}
we have the following sub-matrices \cite{SM}
\begin{align}
\mathcal{J}_{\alpha jk, l}^{AD}&=\mathcal{I}_{\alpha jk, l}^{AD}=0,\label{QFIM2}\\
\mathcal{J}_{k,l}^D&=\frac{M_d}{1+[d]}\mathcal{I}_{k,l}^D=\sum_{m=0}^{d-1}\frac{c_{km}c_{lm}}{4\lambda_m}.\label{QFIM3}\\
\mathcal{J}^A_{\alpha jk,\beta lm}&=M_d\mathcal{I}^A_{\alpha jk,\beta lm}=\frac{1}{\lambda_j+\lambda_k}\delta_{\alpha\beta}\delta_{jl}\delta_{km},\label{QFIM1}
\end{align}
with $c_{km}=\langle m|\sigma_k^D|m\rangle$. Comparing the expressions for the CFIM and the QFIM, we obtain the inequality
\begin{equation}
{\cal I}\ge\frac{1}{2d-1+[d]}{\cal J},
\end{equation}
and with the Gill-Massar lower bound \eqref{conditionIopt}, we get that
\begin{equation}
{\cal I}\ge\alpha_d^{-1}{\cal I}^{(opt)}~~{{\rm or}}~~{\cal I}^{-1}\le\alpha_d[{\cal I}^{(opt)}]^{-1},
\label{MASTER}
\end{equation}
where $\alpha_d=(2d-1+[d])/(d+1)$. Thus, in the worst case, the estimation provided by the set ${\cal B}$ of bases lead to a CFIM that is $\alpha_d^{-1}$ times the optimal one. From Eq.\thinspace(\ref{MASTER}) we obtain ${\cal J}{\cal I}^{-1}\le\alpha_d{\cal J}[{\cal I}^{(opt)}]^{-1}$. Since the condition ${\cal C}={\cal I}^{-1}$ holds, the previous inequality becomes ${\cal J}{\cal C}\le\alpha_d{\cal J}[{\cal I}^{(opt)}]^{-1}$, which together with the relationship between the mean infidelity and the Bures distance lead to
\begin{equation}
\bar I(\rho)\le\alpha_d \bar I^{(opt)}.
\label{MEAN-RESULT}
\end{equation}
Thereby, the mean infidelity provided by the HAQT method here proposed is upper bounded by $\alpha_d$ times the optimal mean infidelity, and lower bounded by the optimal infidelity, that is, $\bar I^{(opt)}\le\bar I(\rho)\le\alpha_d \bar I^{(opt)}$. The proportionality constant $\alpha_d$ becomes 1 for $d=2$, and consequently the optimum $\bar I=\bar I^{(opt)}$ is attained. Since in the limit of large dimensions $\alpha_d$ tends to $2$, we obtain that $\bar I\le2\bar I^{(opt)}$ for larger dimensions.

{\it Experiment.---}  In order to test the HAQT method we employ the transverse momentum of single photons to encode path qudit states. The experimental setup is depicted in Fig~\ref{fig:setup}, and consists of a state preparation (SP) block and a projective measurement (PM) block. At the SP block, the light source consists of a continuous wave (CW) laser operating at $690n$m. The laser is sent to an acousto-optical modulator (AOM) to generate $40n$s wide pulses. Optical attenuators at the output of the AOM set the average number of photons to $\mu=0.1$, so that 90\% of the non-null pulses contain only one photon, giving a non-deterministic single-photon source that is commonly adopted in quantum communications \cite{Gisin,Lo,Diamanti}.
\par
The path qudit is created by defining $d$ possible paths available for the transmission of single-photons through a diffractive aperture \cite{GLima01}. If the transverse coherence length of the beam is larger than the separation between the first and last slit, the state of the transmitted photon is given by \cite{GLima01}
\begin{equation}
|\psi\rangle = \frac{1}{\sqrt{M}}\sum_{l=1}^{d}\sqrt{t_{l}}e^{i\phi_{l}}|l\rangle,
\label{quditspa} \end{equation} where $|l\rangle$ represents the photon state transmitted by the $l$-th slit, $M$ is a normalization constant, and $t_l$ and $\phi_l$ are the transmissivity and relative phase of slit $l$, respectively. All slits are $96 \mu$m wide with $160 \mu$m center to center separation between consecutive slits.
\par
To produce the state and perform the projective measurements required in this protocol, we use a set of ten parallel slits addressed in a sequence of four programmable spatial light modulators (SLM) \cite{GLima03,TomoMubs}. SLMs are optical elements  that can dynamically modulate the amplitude and/or phase of light \cite{Moreno}, and have been a versatile tool for high-dimensional quantum information processing tasks over the last few years \cite{Armin_2018,QRAC1024,GCanas00,MSolis01,Pimenta,LRebon}. In the SP block, the amplitude and phase modulations are obtained with a combination of two SLMs. SLM1 is set for amplitude-only modulation and controls the real part of the state, while SLM2 controls the relative phases \cite{Armin_2018,QRAC1024}. Lenses L2 are used to place the image plane of SLM1 at the position of SLM2.

\begin{figure}[t!]
\centering  	
\includegraphics[width=0.48 \textwidth]{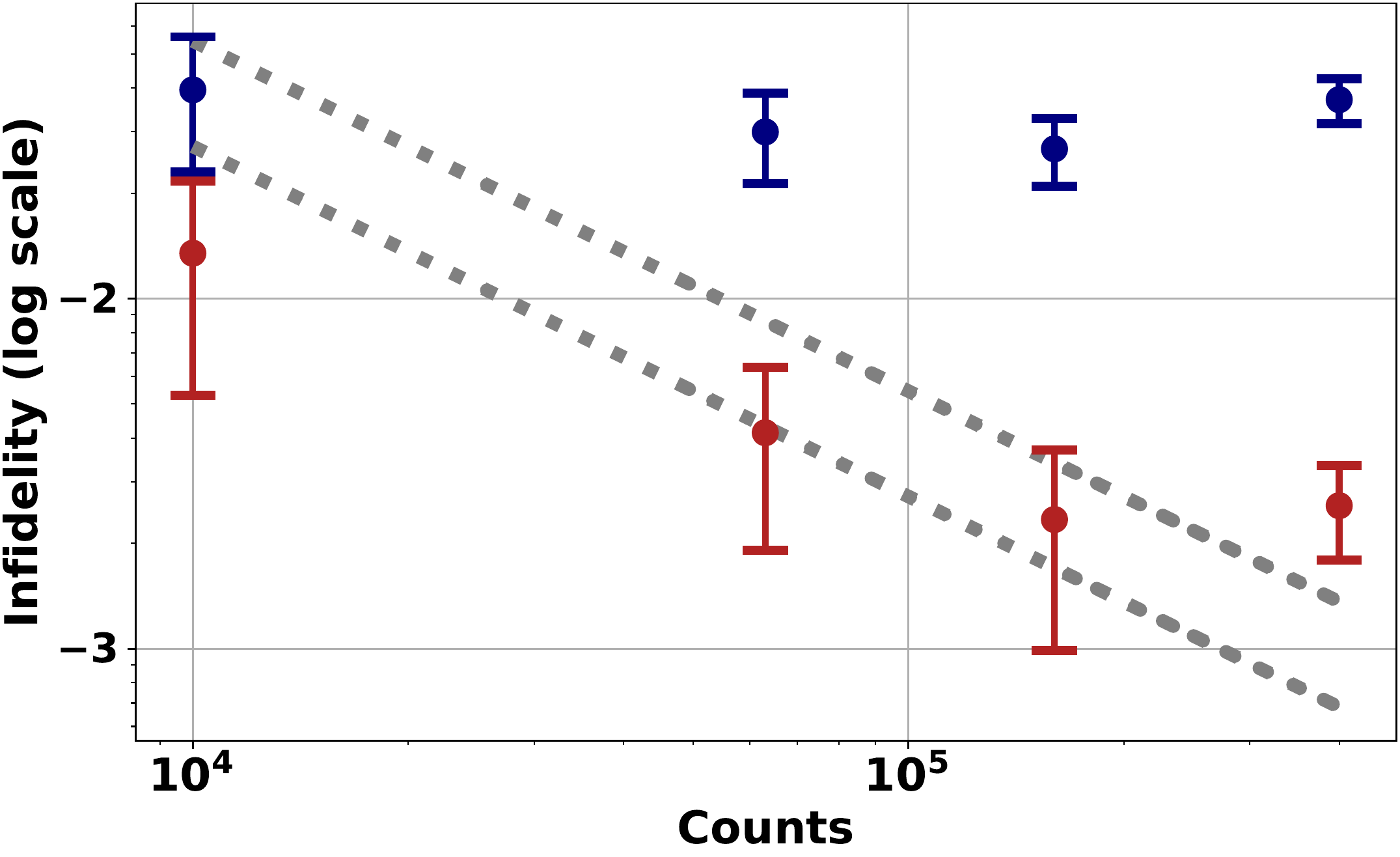}
\caption{Infidelities obtained via SQT (blue dots) and HAQT (red dots) for estimating the state of a 10-dimensional path qudit state, while considering different ensemble sizes. Error bars are calculated considering Poissonian statistics for photon detection. The upper and lower dotted lines represent the upper and lower bounds for the \textit{mean} infidelity, given by $\alpha_{10}\bar I^{(opt)}$ and $\bar I^{(opt)}$, respectively.}
\label{fig:results}
\end{figure}

A $4f$ lens system L3 propagates the state to the PM block, where we use a second pair of SLMs (SLM$3$, SLM$4$) to implement the projective measurement. Similar to the SP block, SLM$3$ (SLM$4$) controls the real (imaginary) part of the projective measurement.  SLM4 is also located at the image plane of SLM3. The measurement is realized with a point-like avalanche photodetector (APD) placed at the center of the focal plane of the last lens L5. The point-like detector is built with a $10\mu$m diameter pinhole placed in front of the APD. In this configuration, the single-photon detection rate is proportional to the overlap between the generated and post-selected states \cite{TomoMubs,GCanas00,GLima09}.

The experimental setup is controlled by two field-programmable gate array (FPGA) electronic units and operates with a repetition rate of $30$Hz. FPGA$1$ controls the SP SLMs and the AOM, while FPGA$2$ controls the PM SLMs and records the number of counts detected by the APD. The synchronization between the two FPGA units allows us to prepare the state and perform a desired projective measurement on each experimental run.

{\it Results.---} To test the new quantum tomography protocol, we prepare a state close to

\begin{equation}
|\Psi\rangle = \frac{1}{\sqrt{10}}\left[ |1\rangle +\sum_{l=2}^{10} e^{-i\pi/10}|l\rangle \right]. \label{targetstat}
\end{equation}  We reconstruct the state using both standard quantum tomography (SQT) and our method, for four ensemble sizes defined by the total number of single-photon counts $N = \{10000, 63000, 158500, 398100\}$ registered in different experimental runs. For each ensemble used, we first perform SQT using $N$ counts and obtain its corresponding estimate using the maximum-likelihood estimation (MLE) technique. Then, to implement our HAQT procedure, we use $N/2$ counts for another round of SQT/MLE to estimate the state $\rho_0$ used to determine the adapted bases. We then measured another $N/2$ photons in these new bases to obtain the final state.

To generate a benchmark for the accuracy of the two methods we follow closely the procedure employed in \cite{Mahler,Huo}, where the total counts acquired by all experimental runs are used to obtain the most accurate estimate of the prepared state.  We then use the infidelity between this high-fidelity estimate and those obtained in each experimental run to compare the two methods.  Fig.~\ref{fig:results} shows the obtained infidelities of SQT and HAQT for estimating the 10-dimensional qudit state for the different values of $N$. The overall behavior of both methods exhibits a decrease of the infidelity for the first three points followed by a slightly increase of the infidelity in the last point. The later indicates the onset of stagnation, where other sources of error such as, for instance, misalignment due to the long run time of the measurement dominate over the error due to finite ensemble size. Fig.~\ref{fig:results} also displays for reference purposes the bounds $\alpha_{10}\bar I^{(opt)}$ and $\bar I^{(opt)}$. The first point obtained by HAQT is below both bounds, which is possible since the bounds are for the mean infidelity, while the next two points are in between. The infidelities obtained by SQT are well above both bounds. Additional numerical simulations of the HAQT method can be found in Ref. \cite{SM}. This clearly indicates that accuracy in state estimation can be significantly improved using our adaptive HAQT method, and this may be particularly important for the complete analysis of high-dimensional quantum systems, processes, and devices.

{\it Conclusions.---}With the advent of integrated photonic quantum technologies \cite{NatPhotReview}, space-division multiplexing optical fibers \cite{GuixReview_2019}, and superconducting circuits \cite{GWendinReview}, there has been a surge of new quantum hardware capable of exploiting advantages provided by high-dimensional quantum systems. Naturally, high-dimensional implementations are more challenging and time-intensive when compared with qubit based experiments and, therefore,  require new tomographic methods that achieve a required precision with a smaller number of measurements.  In this work, we introduce a new tomographic method that is characterized by an accuracy that is close to the Gill-Massar bound for any finite dimension. We demonstrated the improved accuracy of the protocol on a ten-dimensional photonic system.  Since, in general, it is not known whether this bound can be achieved, our work defines a new achievable accuracy limit for the estimation of quantum states.
\begin{acknowledgments}
We thank M. Casanova for his technical assistance. This work was supported by Millennium Institute for Research in Optics (MIRO) and by Fondo Nacional de Desarrollo Científico y Tecnológico Grants 1200859, 1190933, 3200779, 1200266, 1180558 and 72200275.
\end{acknowledgments}

\end{document}